\documentclass[a4paper,10pt,twocolumn,aps,amsmath,amssymb,superscriptaddress,floatfix,nofootOAinbib]{revtex4-1}

\usepackage{graphicx}
\usepackage{dcolumn}
\usepackage{bm}
\usepackage{xcolor}
\usepackage{ulem}
\usepackage{url}
\usepackage{longtable}

\usepackage{subfigure}
\usepackage{multirow}

\usepackage{booktabs}
\usepackage{tabularx}

\begin{document}

\title{Gluon distributions and mass decompositions of the pion and kaon}

\author{Chengdong Han}
\email{chdhan@impcas.ac.cn}
\affiliation{Institute of Modern Physics, Chinese Academy of Sciences, Lanzhou 730000, China}
\affiliation{School of Nuclear Science and Technology, University of Chinese Academy of Sciences, Beijing 100049, China}

\author{Wei Kou}
\email{kouwei@impcas.ac.cn}
\affiliation{Institute of Modern Physics, Chinese Academy of Sciences, Lanzhou 730000, China}
\affiliation{School of Nuclear Science and Technology, University of Chinese Academy of Sciences, Beijing 100049, China}

\author{Rong Wang}
\email{rwang@impcas.ac.cn (corresponding author)}
\affiliation{Institute of Modern Physics, Chinese Academy of Sciences, Lanzhou 730000, China}
\affiliation{School of Nuclear Science and Technology, University of Chinese Academy of Sciences, Beijing 100049, China}

\author{Xurong Chen}
\email{xchen@impcas.ac.cn (corresponding author)}
\affiliation{Institute of Modern Physics, Chinese Academy of Sciences, Lanzhou 730000, China}
\affiliation{School of Nuclear Science and Technology, University of Chinese Academy of Sciences, Beijing 100049, China}

\begin{abstract}
  We present the gluon distribution functions of pion and kaon
  in small-$x$ and large-$x$ regions,
  and compare them with the results obtained from lattice QCD and continuum Schwinger function methods.
  Whether in the small-$x$ region or the large-$x$ region,
  our gluon distribution of pion is consistent with the results of
  lattice QCD and continuum Schwinger function method.
  In addition, the first four moments of gluon distributions of pion and kaon
  at different $Q^{2}$ scales are calculated.
  Furthermore, we present the mass decompositions of pion and kaon
  with the dynamical parton distribution functions calculated
  by the DGLAP equation
  with parton-parton recombination corrections.
  The mass structures of
  pion and kaon are totally different from that of the proton.
\end{abstract}

\maketitle

\section{Introduction}
\label{introduction}
The pion, the lightest bound state in quantum chromodynamics (QCD),
plays an important role in nuclear physics as
it is a Nambu-Goldstone boson \cite{Nambu:1960tm,Goldstone:1961eq}
with dynamical chiral symmetry breaking (DCSB).
For the kaon meson, although it was discovered in the 1950s \cite{Rochester:1947mi},
little is known about its structure.
The studies of the structures
of pion and kaon can reflect the physics of DCSB,
which is helpful to reveal the relative influence of DCSB
on the chiral symmetry breaking of the quark masses,
and it is of great significance to understand the non-perturbative QCD.
Studying the pion and kaon parton distribution functions (PDFs) is
not only for characterizing their structures but also for
further understanding of DCSB and non-perturbative QCD.
At present, people know less about the pion and kaon PDFs than the nucleon PDFs, for
the experimental data are scarce,
especially for the gluon distributions.

Most of the global analyses of pion PDFs mostly rely on Drell-Yan data.
In our previous global QCD analysis \cite{Han:2020vjp},
studies of pion and kaon PDFs were based mostly on pion-induced Drell-Yan data \cite{Saclay-CERN-CollegedeFrance-EcolePoly-Orsay:1980fhh, NA3:1983ejh, NA10:1985ibr, Conway:1989fs},
and the leading-neutron deep inelastic scattering (DIS) data
of $e-p$ collision at HERA \cite{ZEUS:2002gig,H1:2010hym}.
There are some recent studies, including the work of Bourrely and Soffer \cite{Bourrely:2018yck}
that extract the pion parton distributions
from the Drell-Yan $\pi$+$W$ data with quantum statistical approach.
JAM Collaboration \cite{Barry:2018ort,Cao:2021aci} performed
the first Monte Carlo global QCD analysis of
the Drell-Yan $\pi$A data and the leading-neutron DIS data from HERA to reach
the low-$x$ region.
This Monte Carlo global QCD analysis of pion PDFs by the JAM Collaboration reveals
that the gluons carry significantly higher momentum fraction
than that only inferred from Drell-Yan data.
The xFitter developers' team \cite{Novikov:2020snp} used their open source
QCD fitting framework to promote a PDF extraction from the Drell-Yan $\pi$A and photon production data.
It is found that these data
can well constrain the valence distribution,
but they are not sensitive enough to the accurate determination
of sea and gluon distributions.
Chang et al.\cite{Chang:2020rdy} show that the pion-induced $J/\psi$ production data
can impose useful additional constraints
on pion PDFs, especially for the gluon distribution of the pion in the large-$x$ region.

As we know, the studies of pion and kaon gluon distributions
from a theoretical perspective
can provide useful information for experiments.
Most theoretical model calculations only predict the valence quark distribution of pion
\cite{Nam:2012vm,Watanabe:2016lto, Watanabe:2017pvl, Hutauruk:2016sug, deTeramond:2018ecg, Watanabe:2019zny, Han:2018wsw, Chang:2014lva, Chang:2014gga, Chen:2016sno, Shi:2018mcb, Bednar:2018mtf, Ding:2019lwe},
while gluon and sea PDFs are predicted by the Dyson-Schwinger equation (DSE)
continuum method with QCD evolution \cite{Watanabe:2017pvl}.
The description of the pion gluon PDF that based on
the Rainbow-Ladder truncation of DSE is consistent
with the pion gluon PDF results of JAM Collaboration \cite{Barry:2018ort,Cao:2021aci} within two sigma.
Recently, Fan and Lin present the $x$-dependent gluon distribution
of pion from lattice-regularised QCD (LQCD)
with the pseudo-PDF approach \cite{Fan:2021bcr} for the first time.

In this work, we present our $x$-dependent pion gluon distribution
from a global QCD analysis of the available experimental data \cite{Han:2020vjp}
and compare it with the results of LQCD calculations, continuum Schwinger function methods (CSMs),
and other theoretical calculations.
The $x$-dependent gluon distribution of kaon is also presented
under a certain scale $Q^{2}$.
In addition, the first four moments of gluon distributions
of pion and kaon at different $Q^{2}$ scales are calculated and
compared with some model results of DSE \cite{Cui:2020tdf}, JAM \cite{Barry:2018ort,Cao:2021aci},
xFitter \cite{Novikov:2020snp} and LQCD \cite{Salas-Chavira:2021wui} respectively.
Furthermore, according to the hadron mass
decomposition which is
based on the structure
of the QCD energy-momentum tensor and the pion and kaon PDFs
from our global QCD analysis,
we study the mass structures of the pion and kaon.

\section{Gluon Distributions of Pion and Kaon}
\label{SecII:determine initial PDFs}
\begin{figure}[htp]
\begin{center}
\includegraphics[width=0.45\textwidth]{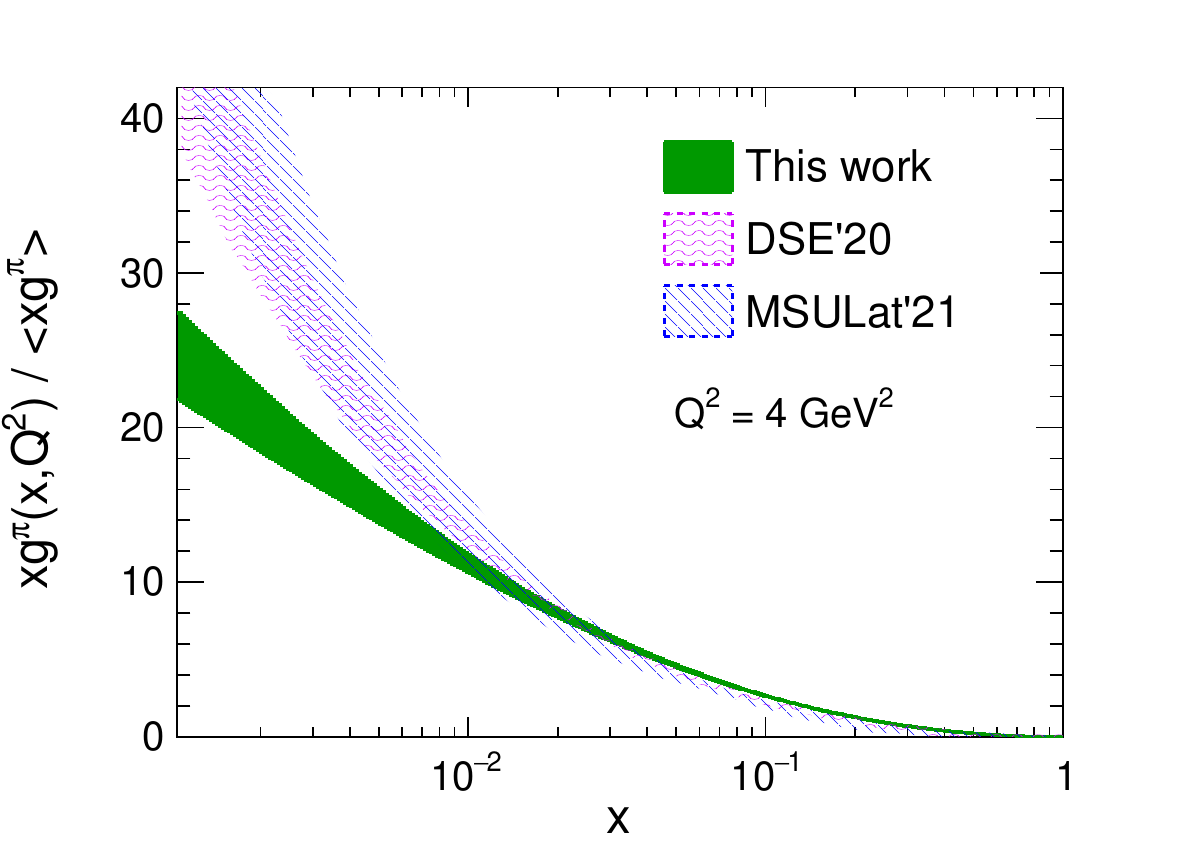}
\includegraphics[width=0.45\textwidth]{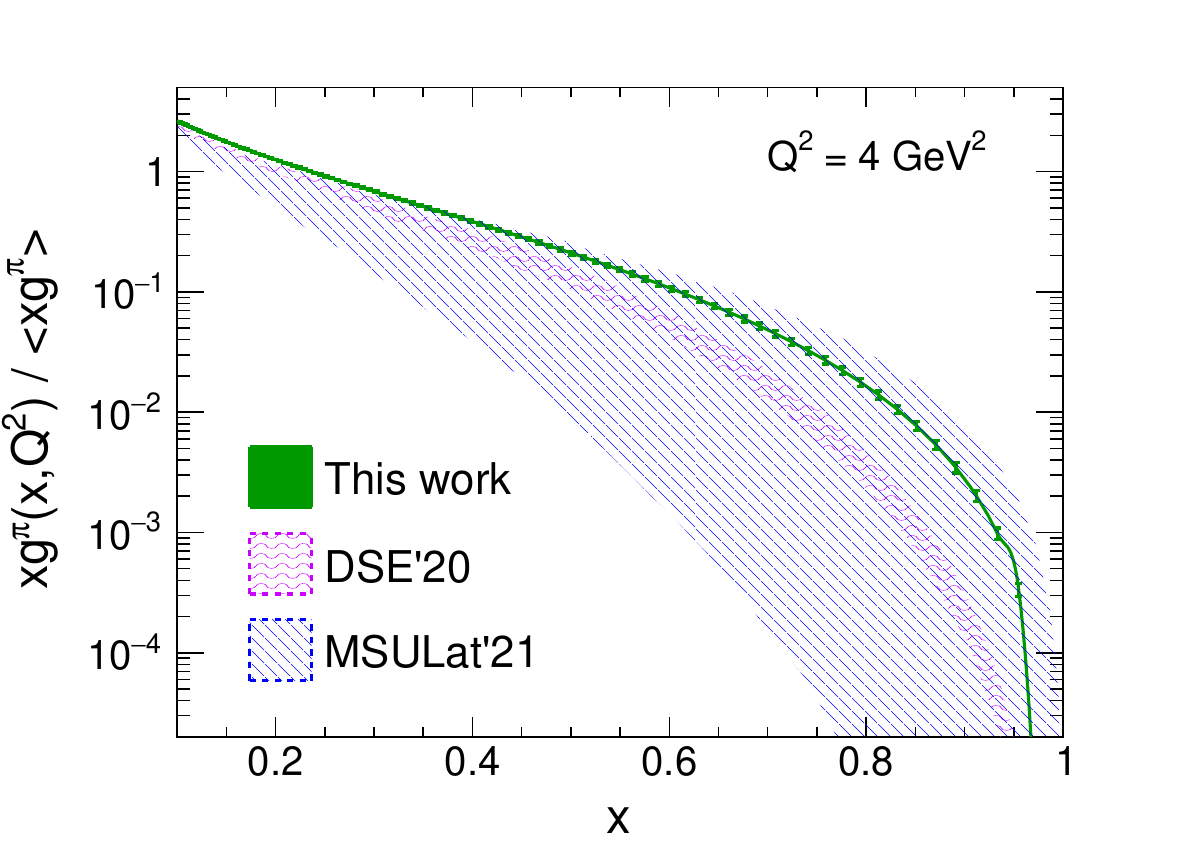}
\caption{~(Color online)~
  (Up)~Comparisons of our result to the CSMs \cite{Cui:2020dlm, Cui:2020tdf}
  and LQCD \cite{Fan:2021bcr} calculation results for
  the pion gluon distribution in the small-$x$ region.
  ~(Bottom)~Comparisons of our result to the CSMs \cite{Cui:2020dlm, Cui:2020tdf}
  and LQCD \cite{Cui:2020dlm, Cui:2020tdf} for the pion
  gluon distribution in the large-$x$ region.
}
\label{xgluon_pion}
\end{center}
\end{figure}

Fan and Lin recently showed the determination of
the $x$-dependent pion gluon distribution from LQCD
using the pseudo-PDF approach \cite{Fan:2021bcr}.
Besides, Cui et al. predicted the pion's gluon distribution
with CSMs \cite{Cui:2020tdf}.
In this work, we present pion and kaon PDFs with uncertainties
from a global QCD analysis of the experimental data
within the framework of dynamical parton model \cite{Han:2020vjp}.
Figure \ref{xgluon_pion} displays the normalized pion gluon distribution
$xg^{\pi}(x,Q^{2})/\left<xg^{\pi}\right>$ at $Q^{2}$ = 4 GeV$^{2}$ from us,
compared with the LQCD \cite{Fan:2021bcr} and DSE \cite{Cui:2020tdf,Cui:2020dlm} calculations.
The top pad
in Fig.\ref{xgluon_pion} shows the gluon distribution in the small-$x$ region,
and the bottom pad
in Fig.\ref{xgluon_pion} shows the gluon distribution in the large-$x$ region.
There shows agreements among the three results of gluon distribution
on the entire depicted domain. Judged by the agreement
between our work and other theoretical predictions (LQCD and CSMs)
for the pion gluon distribution in Fig.\ref{xgluon_pion},
our pion gluon distribution obtained from global QCD analysis is also reliable.

\begin{figure}[htp]
\begin{center}
\includegraphics[width=0.45\textwidth]{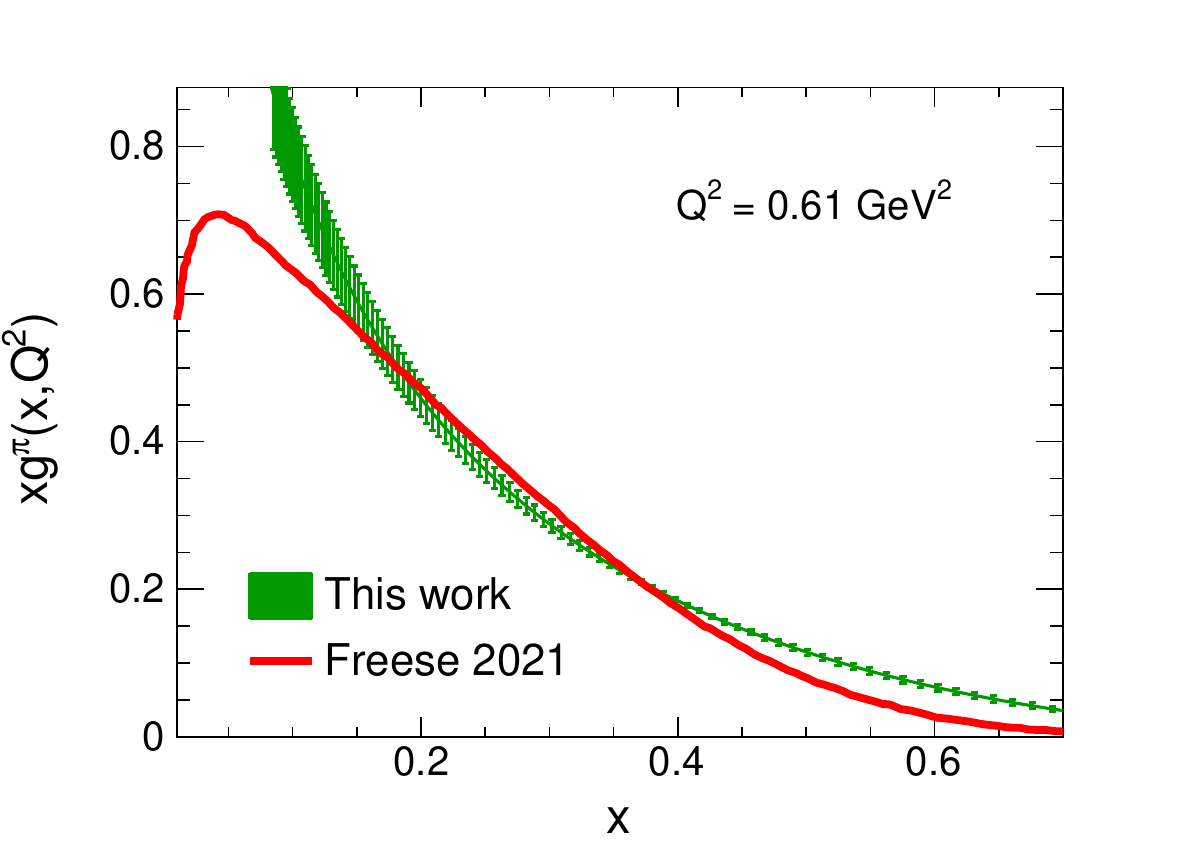}
\caption{~(Color online)~
 Comparisons of our result to the gluon PDF from quark dressing
 in the pion \cite{Freese:2021zne} at Q$^{2}$ = 0.61 GeV$^{2}$.
}
\label{xgluon_Tandy}
\end{center}
\end{figure}
The pion gluon distribution $xg^{\pi}(x, Q^{2})$ we calculated
at $Q^{2}$ = 0.61 GeV$^{2}$ is shown in Fig. \ref{xgluon_Tandy},
compared with the gluon distribution from the dressed quark \cite{Freese:2021zne} in the pion.
By comparison, one sees that our result based on the dynamical parton model is consistent with Freese's result.
The dressing-gluon is derived from the constituent quarks,
thus when $x\textless 0.1$, the distribution of dressing-gluon decreases.
For our result, most of the gluons generated under the dynamical parton model are
from parton splitting process, thus when $x\textless 0.1$,
our dynamical parton model
still has the gluons which are not related to the dressed quarks.
\begin{figure}[htp]
\begin{center}
\includegraphics[width=0.45\textwidth]{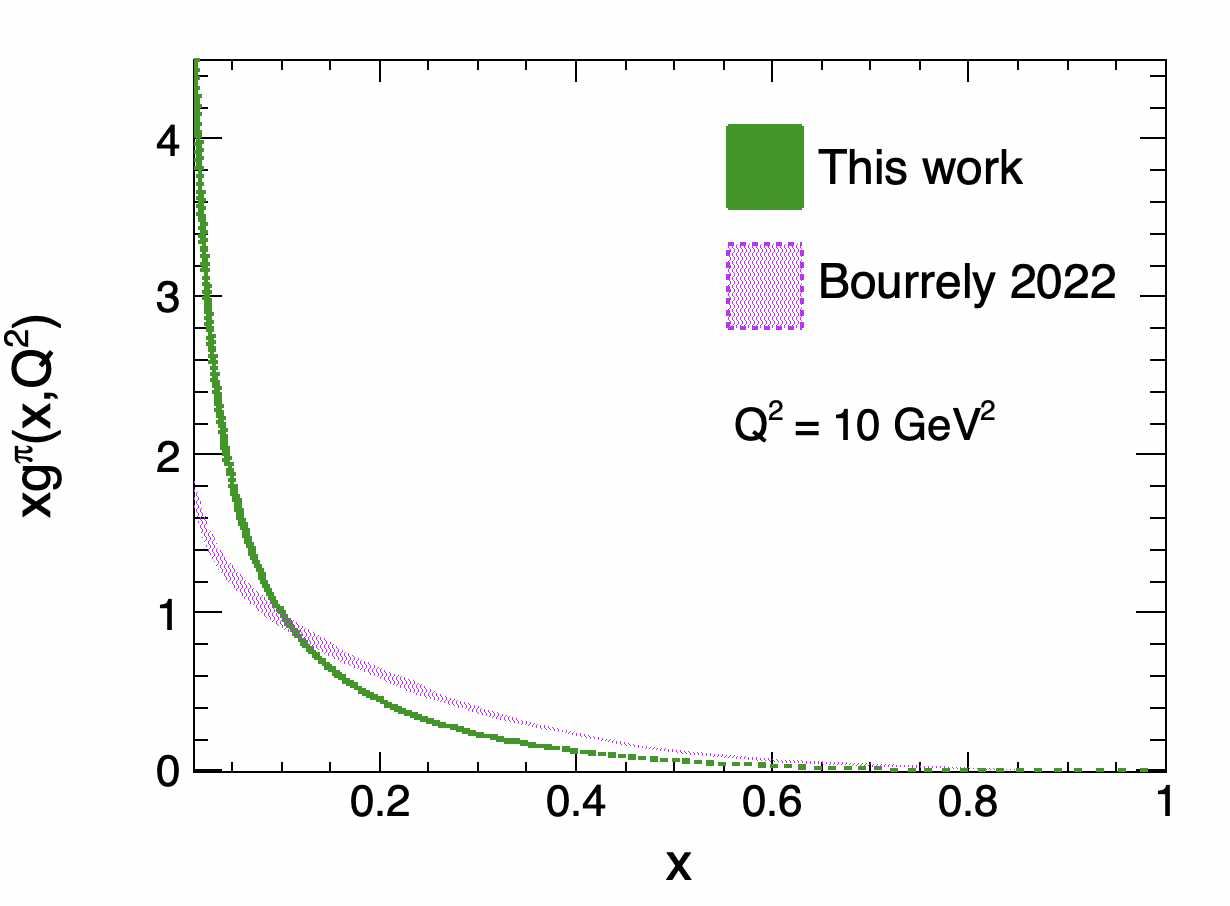}
\caption{~(Color online)~
  Our obtained pion gluon distribution from a global QCD analysis
  compared with the result from a statistical model constrained by
  pion-induced Drell-Yan data and J/$\psi$ production data
  at $Q^{2}$ = 10 GeV$^{2}$ \cite{Bourrely:2022mjf}.
}
\label{xgluon_Bourrely}
\end{center}
\end{figure}
Figure \ref{xgluon_Bourrely} shows our pion gluon distribution $xg^{\pi}(x, Q^{2})$
at $Q^{2}$ = 10 GeV$^{2}$, compared to the pion gluon distribution from the framework
of a statistical model fitted to the
pion-induced Drell-Yan data and J/$\psi$ production data of the pion \cite{Bourrely:2022mjf}.

\begin{figure}[htp]
\begin{center}
\includegraphics[width=0.45\textwidth]{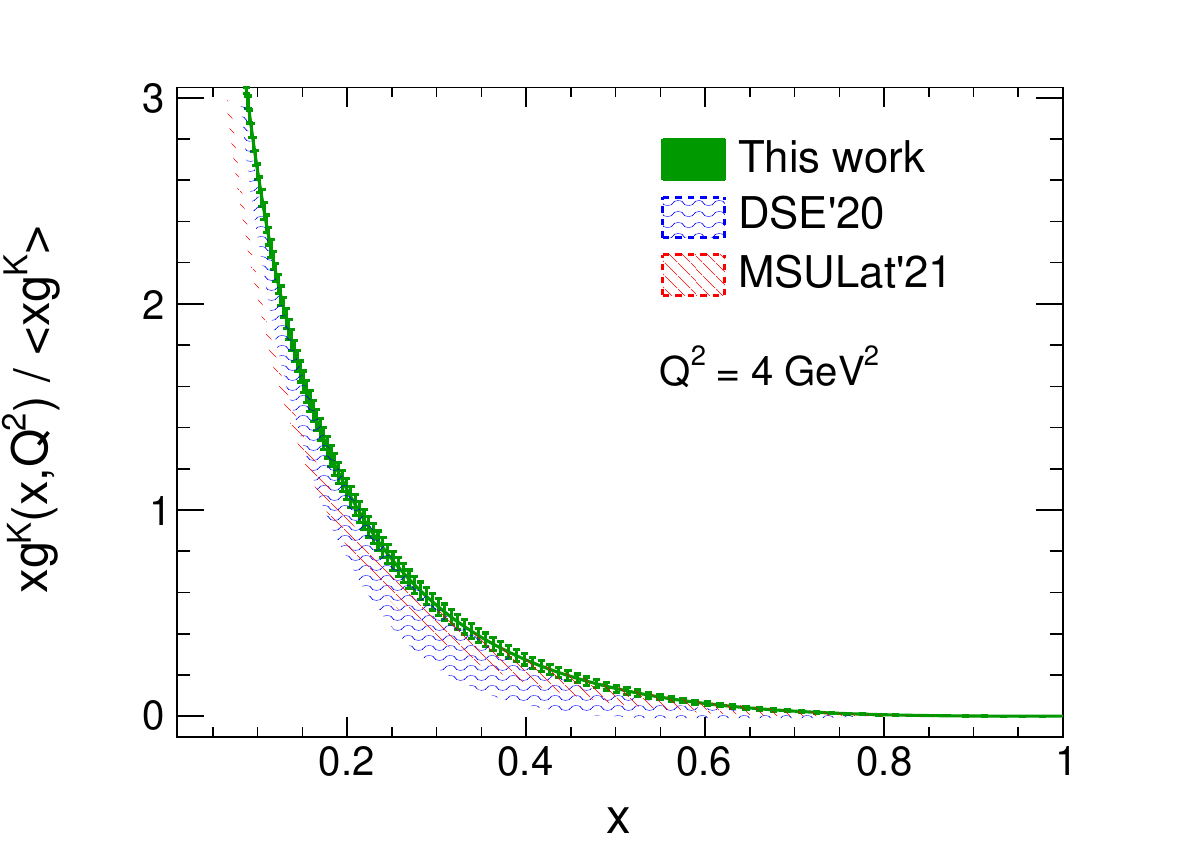}
\caption{~(Color online)~
  Comparisons of our kaon gluon distribution with the DSE \cite{Cui:2020tdf} 
  and LQCD \cite{Salas-Chavira:2021wui} results in the large-$x$ region.
}
\label{xgluon_kaon}
\end{center}
\end{figure}

Figure \ref{xgluon_kaon} shows our normalized gluon distribution 
$xg^{K}(x,Q^{2})/\left<xg^{K}\right>$ of the kaon 
in the large-$x$ region at $Q^{2}$ = 4 GeV$^{2}$.
By comparing the kaon gluon distributions, one sees 
that our result is consistent with the DSE'20 \cite{Cui:2020tdf} in the $\overline{MS}$ scheme 
and the lattice QCD calculation \cite{Salas-Chavira:2021wui}.
\begin{figure}[htp]
\begin{center}
\includegraphics[width=0.45\textwidth]{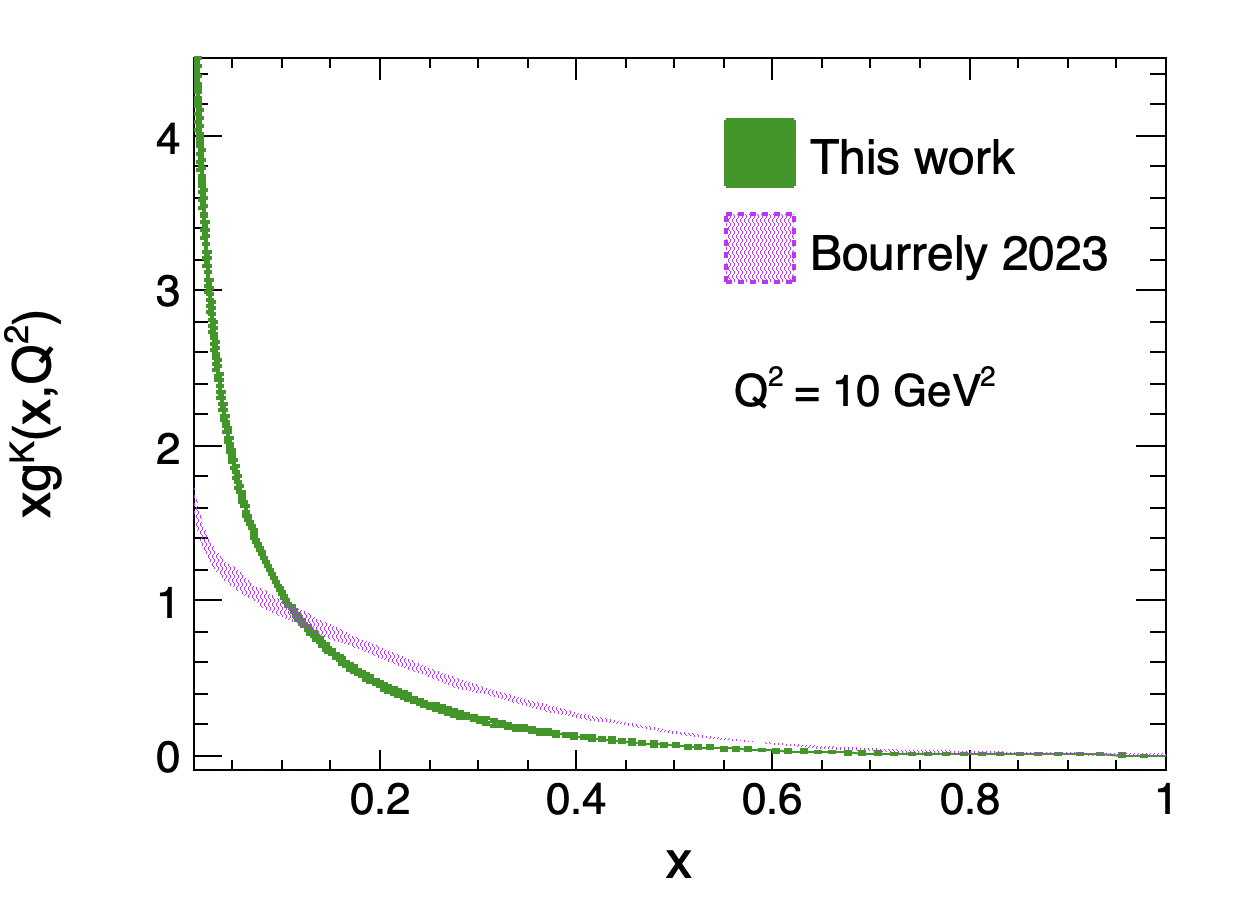}
\caption{~(Color online)~
  Our kaon gluon distribution from a global QCD analysis 
  compared with Bourrely's result from a statistical model
  constrained by the meson-induced Drell-Yan and quarkonium production data 
  at $Q^{2}$ = 10 GeV$^{2}$ \cite{Bourrely:2023yzi}.
}
\label{xgluon_Bouurely_kaon}
\end{center}
\end{figure}
Our kaon gluon distribution $xg^{K}(x, Q^{2})$ at $Q^{2}$ = 10 GeV$^{2}$ from a global QCD analysis is shown in Fig. \ref{xgluon_Bouurely_kaon},
compared with Bourrely's result 
from an analysis of meson-induced Drell-Yan and quarkonium production data 
within a statistical model \cite{Bourrely:2023yzi}.

In order to compare with more other theoretical calculations,
we calculate the first four moments of the pion and kaon gluon distributions at different Q$^{2}$.
The moments of the 
gluon distribution is defined as follows,
\begin{equation}
<x^{n} g>= \int_0^1 x^{n} g(x,Q^2) dx, 
\label{MomentumSum}
\end{equation}
where $g(x,Q^{2})$ is the gluon distribution.

\begin{figure*}[htp]
\begin{center}
\includegraphics[width=0.85\textwidth]{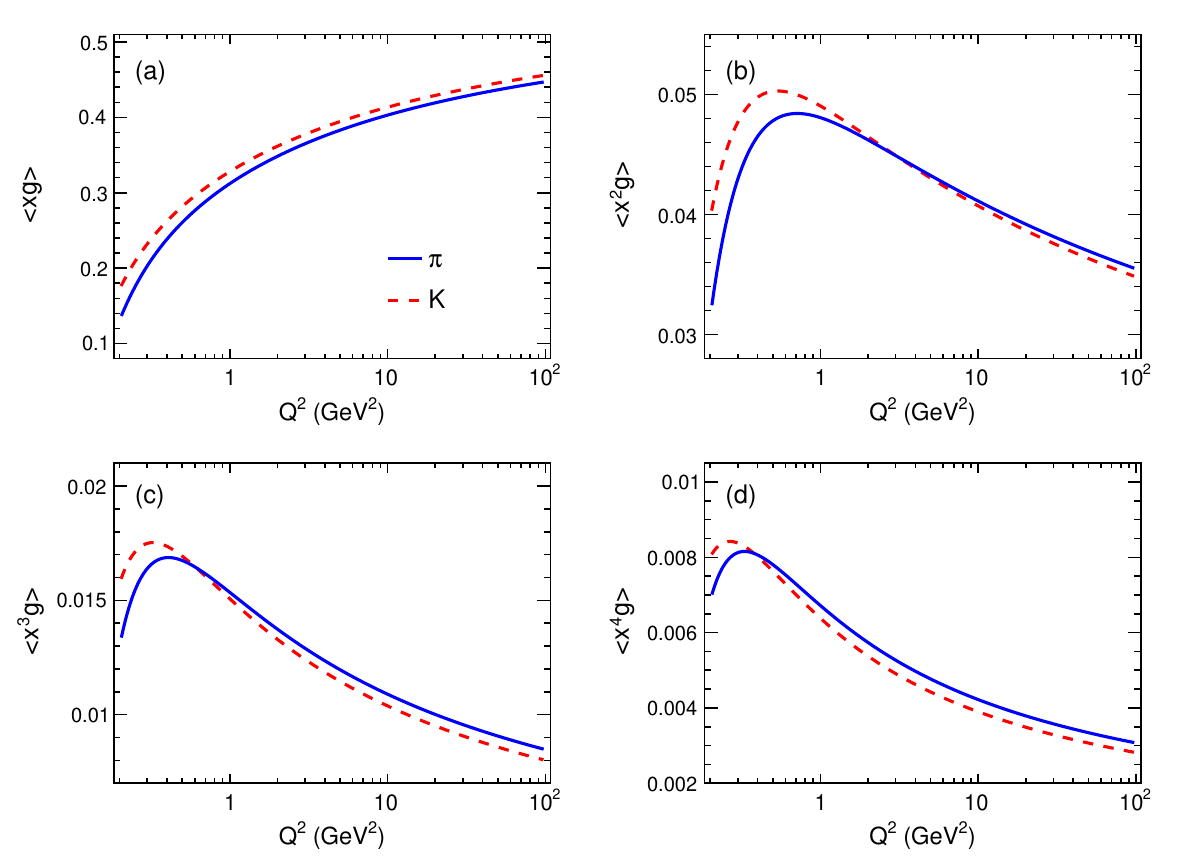}
\caption{~(Color online)~
  The first four moments of the pion and kaon gluon distribution functions 
  as a function of the square of four momentum transfer $Q^{2}$. 
  The (a), (b), (c), and (d) panels show the first-order moment, second-order moment, 
  third-order moment, and fourth-order moment, respectively.
  The blue solid and red dashed curves show the first four moments 
  of pion and kaon gluon distributions, respectively.
}
\label{pion_kaon_mom_vs_Q2}
\end{center}
\end{figure*}

Figure \ref{pion_kaon_mom_vs_Q2} shows the first four moments 
of pion and kaon gluon distributions as a function of $Q^{2}$ scale. 
The blue solid and red dashed curves show the evolutions of
the first four moments of pion and kaon gluon distributions, respectively.
The (a), (b), (c), and (d) pads present 
the first-order moment, second-order moment, third-order moment, 
and fourth-order, respectively. 
\begin{table}[h]
  \vspace{-0.18 cm}
  \centering
  \caption{Comparisons of our result from a global QCD analysis 
  in the dynamical parton model with some other model calculations
  for $\left<x\right>_{g}^{\pi}$ and 
  $\left<x\right>_{g}^{K}$ at $Q^{2}$ = 4, 27 GeV$^{2}$. }
  \label{Table_xg}
  	\resizebox{\columnwidth}{!}{
  \begin{tabular}{lll|lll}
    \hline
    Model           & $\left<x\right>_{g}^{\pi}$ & $Q^{2} / GeV^{2}$   & Model    & $\left<x\right>_{g}^{K}$  & $Q^{2} / GeV^{2}$ \\
    \hline
    DSE-RL\cite{Freese:2021zne}  & 0.34            & 4                   & DSE\cite{Cui:2020tdf}    & 0.44           & 27 \\
    xFitter\cite{Novikov:2020snp}& 0.25            & 4                   & This work                 & 0.39            & 4  \\
    GRV\cite{Gluck:1991ey}    & 0.51               & 4                   & This work                 & 0.44            & 27 \\
    Ref.\cite{Ding:2019lwe}    & 0.41              & 4                   &          &                & \\
    LQCD\cite{Shanahan:2018pib}& 0.61              & 4                   &          &                & \\
    DSE\cite{Cui:2020tdf}     & 0.41               & 4                   &          &                & \\
    This work                  & 0.38              & 4                   &          &                & \\
    \hline
  \end{tabular}}
  \vspace{-0.4 cm}
\end{table}
The first moments of the gluon distributions of pion and kaon are listed in Table \ref{Table_xg}.
Our $\left<x\right>_{g}^{\pi}$ and $\left<x\right>_{g}^{K}$ predictions agree well with the other model calculations.
It is worth noting that LQCD and DSE are two main nonperturbative approaches derived directly from the QCD theory.
\begin{table}[h]
  \vspace{-0.18 cm}
  \centering
  \caption{Comparisons of our result from a global QCD analysis 
    in the dynamical parton model with some other model calculations
    for $\left<x^{2}\right>_{g} / \left<x\right>_{g}$, 
    $\left<x^{3}\right>_{g} / \left<x\right>_{g}$ of pion 
    and kaon at $Q^{2}$ = 4 GeV$^{2}$. }
  \label{Table_xng_xg}
  	\resizebox{\columnwidth}{!}{
  \begin{tabular}{ll|ll}
    \hline
    Model      & $\left<x^{2}\right>_{g}^{\pi} / \left<x\right>_{g}^{\pi}$  & Model   
               & $\left<x^{2}\right>_{g}^{K} / \left<x\right>_{g}^{K}$  \\
    \hline
    DSE \cite{Cui:2020tdf}              & 0.076                                                     & DSE \cite{Cui:2020tdf}     & 0.075              \\
    JAM \cite{Barry:2018ort,Cao:2021aci}& 0.103                                                     & LQCD \cite{Salas-Chavira:2021wui} & 0.078       \\
    xFitter \cite{Novikov:2020snp}      & 0.158                                                     & This work                  & 0.087              \\
    LQCD \cite{Salas-Chavira:2021wui}   & 0.092                                                     &                            &                    \\
    This work                           & 0.117                                                     &                            &                    \\
    \hline
    Model                        & $\left<x^{3}\right>_{g}^{\pi} / \left<x\right>_{g}^{\pi}$                
        & Model                  & $\left<x^{3}\right>_{g}^{K} / \left<x\right>_{g}^{K}$ \\
    \hline
    DSE \cite{Cui:2020tdf}              & 0.015                                                     & DSE \cite{Cui:2020tdf}     & 0.015             \\
    JAM \cite{Barry:2018ort,Cao:2021aci}& 0.024                                                     & LQCD\cite{Salas-Chavira:2021wui} & 0.019             \\
    LQCD \cite{Salas-Chavira:2021wui}   & 0.048                                                     & This work                  & 0.021             \\
    This work                           & 0.033                                                     &                            &                   \\
    \hline
  \end{tabular}}
  \vspace{-0.4 cm}
\end{table}
Table \ref{Table_xng_xg} lists our 
{\color{blue} predictions of} $\left<x^{2}\right>_{g} / \left<x\right>_{g}$, 
$\left<x^{3}\right>_{g} / \left<x\right>_{g}$ for pion and kaon, which are 
in good agreements with the corresponding results from DSE \cite{Cui:2020tdf}, 
JAM \cite{Barry:2018ort,Cao:2021aci}, xFitter \cite{Novikov:2020snp} 
and LQCD \cite{Salas-Chavira:2021wui} approaches.

\section{Mass structures of the pion and kaon}
\label{Mass_structure_of_the_pion}
The mass structure of pion is very interesting. 
According to Goldstone theorem, the pion is essentially different 
from other ordinary hadrons, which is a collective mode in QCD vacuum.
We will see that the mass structure does reflect that. 
The QCD structure of hadron mass can be decomposed into four different parts:
\begin{equation}
  \begin{aligned}
    M = M_{q} + M_{g} + M_{m} + M_{a}, \\
  \end{aligned}
  \label{mass_sum}
\end{equation}
in which $M_{q}$ is the quark energy contribution, 
$M_{g}$ is the gluon energy contribution, 
$M_{m}$ is the quark mass contribution
and $M_{a}$ is the trace anomaly contribution.

\begin{equation}
  \begin{aligned}
    M_{q} = \frac{3}{4} (a - \frac{b}{1+\gamma_{m}})M, \\
    M_{g} = \frac{3}{4} (1 - a)M, \\
    M_{m} = \frac{4 + \gamma_{m}}{4(1+\gamma_{m})}bM, \\
    M_{a} = \frac{1}{4}(1 - b)M.
  \end{aligned}
  \label{mass_decomposition}
\end{equation}

These four mass terms mainly depend on the momentum fraction carried by all the quarks from our analysis \cite{Han:2020vjp}, 
QCD trace anomaly parameter $b$ calculated via chiral perturbation theory \cite{Ji:1995sv} 
and the quark mass anomalous dimension $\gamma_{m}$. The quark mass anomalous dimension 
$\gamma_{m}$ \cite{Baikov:2014qja} is predicted by perturbative QCD as follows:
\begin{equation}
  \begin{aligned}
    \gamma_{m} = \sum_{i=0}^N(\gamma_{m})_{i}a_{s}^{i+1}, \\
  \end{aligned}
  \label{Gamma_m}
\end{equation}
where $a_{s} = \alpha_{s} / \pi = g^{2}/(4\pi^{2})$, 
and $g$ denotes the renormalized strong coupling constant. 
It is worth noticing that the anomalous dimension $\gamma_{m}$ 
defined by Ji \cite{Ji:1995sv} is -2 times \cite{Wang:2019mza} 
that defined in Ref. \cite{Baikov:2014qja}.
For the running coupling constant $\alpha_{s}$, 
we adopted Cui's latest parametrization 
of the effective charge from lattice QCD \cite{Cui:2019dwv}.
The adopted running coupling constant $\alpha_{s}$ including the gluon mass effect of 
dynamical chiral symmetry breaking is written as:
\begin{equation}
\begin{aligned}
\alpha_{s}(Q^{2}) = \frac{4\pi}{\beta_{0}ln[(m_{\alpha}^{2} + Q^{2}) / \Lambda^{2}_{QCD}]}
\end{aligned}
\label{alpha_s}
\end{equation}

In general, the matrix element $a(Q^{2})$ in Eq. (\ref{mass_decomposition}) 
can be deduced 
from the quark distributions that are measured in $\pi$-N Drell-Yan process.
The matrix element $a(Q^{2})$ is the momentum faction carried 
by the quarks in the pion, which is given by, 
\begin{equation}
  \begin{aligned}
    a(Q^{2}) = \sum_{f} \int_{0}^{1} x[q_{f}(x,Q^{2}) + \bar{q_{f}}(x,Q^{2})].
  \end{aligned}
  \label{a_sum}
\end{equation}
Here $q_{f}(x,Q^{2})$ and $\bar{q_{f}}(x,Q^{2})$ are the quark and anti-quark 
distributions with flavor $f$ in the pion. 
In this work, we use the pion PDFs from our recent global QCD analysis 
of the experimental data to evaluate the matrix element $a(Q^{2})$. 

For the trace anomaly parameter $b$ in Eq. (\ref{mass_decomposition}), 
it can be derived from the pion mass in the chiral perturbation theory.
The first-order chiral perturbation theory gives a clean prediction 
of the trace anomaly parameter to be $b_{\pi}$ = 0.5 \cite{Ji:1995sv}.
\begin{figure}[htp]
\begin{center}
\includegraphics[width=0.45\textwidth]{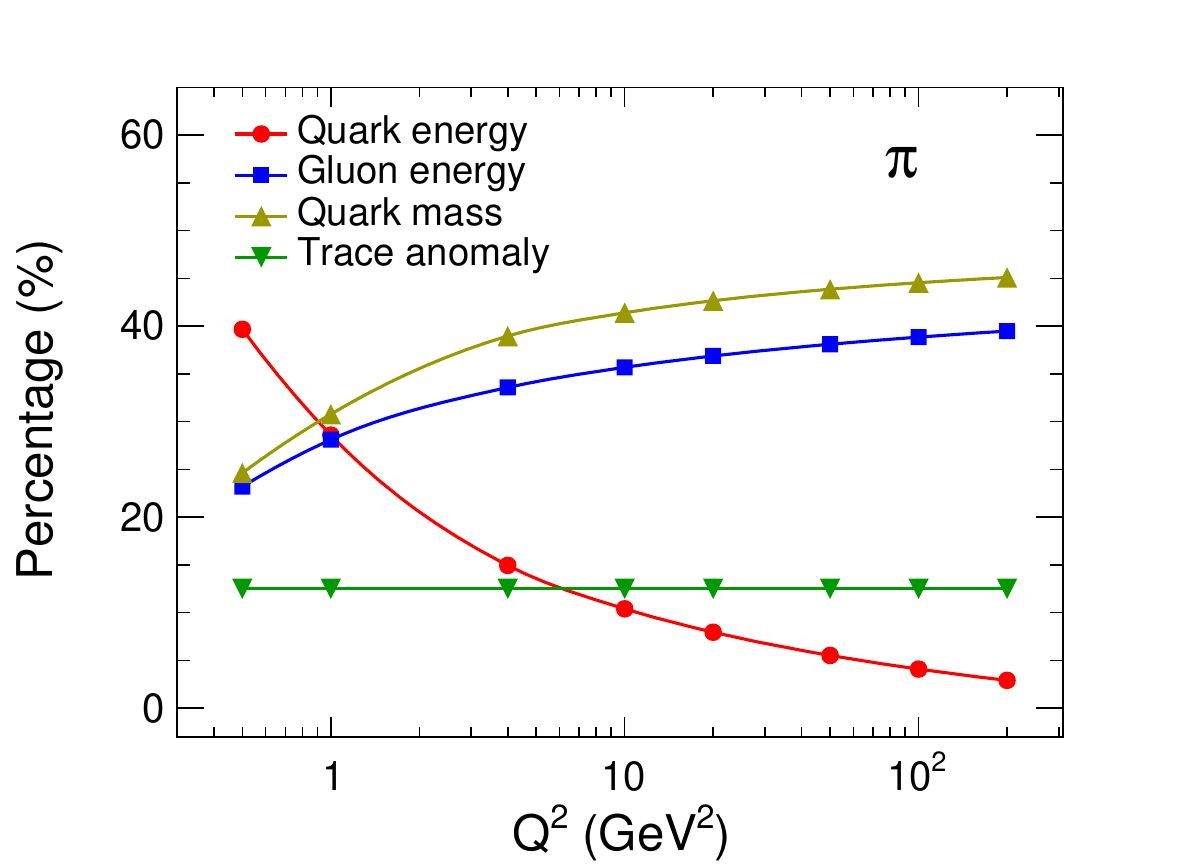}
\caption{~(Color online)~
  The QCD decompositions of the pion mass under different $Q^{2}$ scales, 
  where the different markers indicate the percentages of 
  different contributions to the mass as a function of $Q^{2}$.
}
\label{pion_pie}
\end{center}
\end{figure}

Here we provide a brief explanation 
of the criteria for estimating the trace anomaly parameter $b$. 
First, according to the original derivation of mass decomposition 
and the requirement of Lorentz covariance \cite{Ji:1995sv}, 
$b$ is viewed to be a scale-invariant quantity 
which has been discussed in many previous studies 
\cite{Ji:1995sv,Wang:2019mza,Kou:2021bez,Duran:2022xag,Kou:2023zko}.
Two estimates of trace anomaly parameter $b$ of proton 
in the limits of chiral SU(3) and heavy strange quark 
are $b_{p}$ = 0.17 and $b_{p}$ = 0.11 provided by Ji \cite{Ji:1995sv}. 
Wang et al. firstly present an extraction 
of the QCD trace anomaly parameter 
to be $b_{p}$ = 0.07 $\pm$ 0.17 \cite{Wang:2019mza} from the experimental data 
of near-threshold $J/\psi$ photoproduction cross section from GlueX at JLab. 
Recently, Kou et al. analyze the updated differential cross-section data 
from Hall C and GlueX Collaborations \cite{Duran:2022xag,GlueX:2023pev}, 
and obtain a new value of trace anomaly parameter $b_{p}$ = 0.52 $\pm$ 0.09 \cite{Kou:2023zko},
which is close to the trace anomaly parameter for the pion 
estimated in the chiral perturbation theory \cite{Gasser:1982ap}. 
But this does not necessarily prove that $b$ is independent of the hadrons. 
Second, the mass decomposition of different hadrons should be different. 
Therefore, the trace anomaly parameter $b$ could be different for different hadrons. 
{\color{blue} Third,} we assume that in the chiral limit, 
the first-order chiral perturbation theory under SU(3) gives 
the same mass formula 
for the pion and the kaon. 
This is due to the fact that the vacuum condensations of quarks 
with three flavors are treated equally \cite{Gasser:1982ap}.
Therefore the parameter $b$ 
of the kaon is also taken to be $b_{K}$ = 0.5 in this work.

\begin{figure}[htp]
\begin{center}
\includegraphics[width=0.45\textwidth]{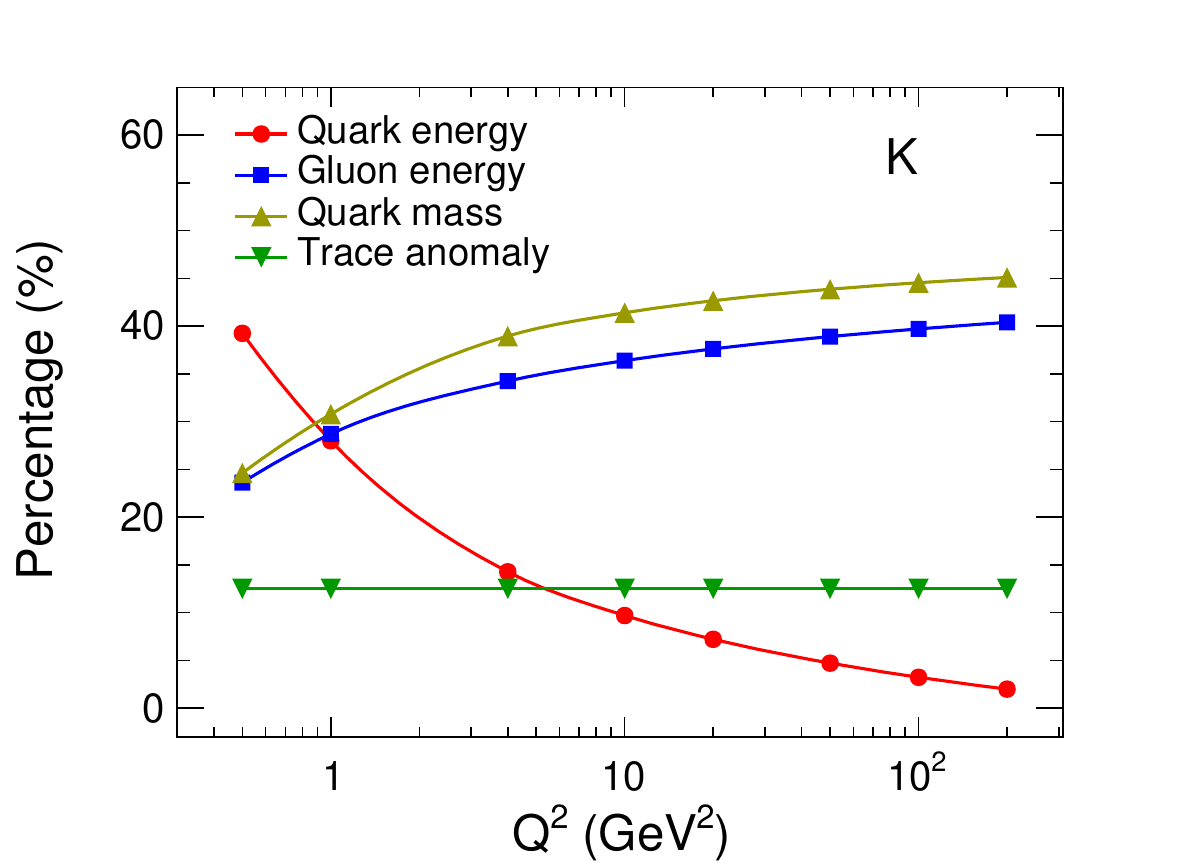}
\caption{~(Color online)~
  The decompositions of the kaon mass under different $Q^{2}$ scales, 
  where the different markers indicate the percentages of 
  different contributions to the kaon mass.
}
\label{kaon_pie}
\end{center}
\end{figure}

Using the trace anomaly parameter $b$ explained above, we calculated the mass decompositions 
of pion and kaon at different $Q^{2}$ scales.
Figure \ref{pion_pie} shows the pion mass decomposition as a function of $Q^2$,
where the different markers indicate the percentages of different contributions to the pion mass. 
By comparing to the mass decomposition of the proton \cite{Wang:2019mza}, 
it is found that the proportion of 
the quark mass contribution in the pion is smaller than that in the proton, 
while the proportion of the trace anomaly contribution in the pion is larger than that in the proton. 
The reason for the large difference between the mass decompositions 
of the pion and the proton is that the trace anomaly parameter $b_{p}$ $\approx$ 0.1
of the proton \cite{Wang:2019mza, Ji:1995sv} is much smaller than that of the pion. 
Quark mass plays a more important role in the pion mass comparing to the proton, 
as the chiral-limit mass of 
the pion is zero while the chiral-limit mass of the proton is huge. 
Moreover, 
the proportion of quark energy $M_{q}$ in the pion is smaller than that in the proton, 
especially at high $Q^{2}$ scale. 
The quark energy $M_{q}$ is the sum of kinetic energy and potential energy.
The reason for the smaller proportion of $M_{q}$ 
in the pion is that the quark potential energy in the pion is lower 
compared to that in the proton. 
In addition to the small proportion of $M_{a}$ to the pion mass, 
the small proportion of $M_{q}$ may be another reason why the pion mass is so light. 
Figure \ref{kaon_pie} shows the 
kaon mass decompositions at different $Q^{2}$ scales, 
where the different markers indicate the percentages 
of different contributions to the kaon mass.

The $Q^2$-dependences of different mass origins in the pion 
or kaon mass decomposition are discussed. 
Firstly, we emphasize that the trace anomaly is scale-invariant, 
which is governed by the symmetry of QCD energy-momentum tensor. 
In other words, the trace anomaly 
parameter $b$ is scale-invariant and it is associated 
with the hadron's scalar charge \cite{Ji:1995sv}.
Secondly, the quark mass term $M_{m}$ depends on the anomalous dimension $\gamma_m$ 
and this gives the $Q^2$-dependence of $M_{m}$. 
Thirdly, the $Q^2$-dependence of the quark energy term $M_{q}$ depends on 
the momentum fraction carried by all quarks $a(Q^2)$, 
and the traditional DGLAP evolution tells us that $a(Q^2)$ decreases as $Q^2$ increases. 
This is the main reason 
why the quark energy term $M_q$ is negatively correlated with the $Q^2$ scale. 
Fourthly, the gluon energy term $M_g$ increases with increasing $Q^2$, and it is also governed by the DGLAP evolution equation. 
With the increasing $Q^2$, the parton splitting processes radiate 
more and more gluons, resulting in a decrease of quark kinetic energy, 
and an increase of gluon kinetic energy, which is in Fig. \ref{pion_kaon_mom_vs_Q2}.
In whole description of the hadron mass decomposition, 
the anomalous dimension $\gamma_m$ only affects the terms related to quarks, 
as the gluon is massless. The gluon contributions to the hadron mass 
only depends on the trace anomaly and the momentum fraction carried by the gluons.

\section{Summary}
\label{summary}
The pion and kaon gluon distributions from our previous 
global QCD analysis applying DGLAP equations with parton recombination
corrections are basically consistent with the DSE and LQCD calculations 
in both small-$x$ and large-$x$ regions. 
The future experimental measurements will provide excellent chances 
to verify these predictions. 
Additionally, the first four moments of the gluon distributions of pion and kaon 
at different $Q^{2}$ scales are predicted,
which agree well with some theoretical models. 

We also give the mass structures of pion and kaon at different $Q^{2}$ scales, 
according to the mass decomposition of QCD energy-momentum tensor.  
The mass of the pion or the kaon can be decomposed into four parts in this work. 
For the pion and the kaon, the quark mass contribution $M_{m}$ 
is the largest one contributed to the meson mass at high $Q^{2}$.
Compared to the proton, the trace anomaly contribution of pion or kaon is small. 
These four origins of the hadron mass depend on the momentum fraction $a$ 
by the quarks, the trace anomaly parameter $b$, and the quark mass anomalous dimension $\gamma_m$. 
In this work, we find some apparent differences between the mass decompositions 
of the pion and the proton \cite{Wang:2019mza}. 
Besides the trace anomaly contribution $M_{a}$, 
we notice that the quark energy contribution $M_{q}$, 
the gluon energy contribution $M_{g}$, and the quark mass contribution $M_{m}$ all have obvious dependence on $Q^{2}$.

The US Electron Ion Collider (EIC) \cite{Aschenauer:2014twa, Aguilar:2019teb, Arrington:2021biu} 
under the construction and the proposed Electron-Ion Collider in China (EicC) 
\cite{Anderle:2021wcy, Chen:2018wyz, Chen:2020ijn} 
provide great opportunities to deepen our understanding on the meson structures 
through the measurements of Sullivan
process \cite{Aguilar:2019teb, Wang:2023thl, Xie:2021ypc}. 
These high-energy projects will surely increase our knowledge 
about the pion and kaon gluon distributions and the emergence of the meson mass.

\begin{acknowledgments}
This work is supported by the National Natural Science Foundation of China under the Grant No. 12305127,
the International Partnership Program of the Chinese Academy of Sciences under the Grant No. 016GJHZ2022054FN,
the Strategic Priority Research Program of Chinese Academy of Sciences under Grant No. XDB34030301 and
the Guangdong Major Project of Basic and Applied Basic Research No. 2020B0301030008.
\end{acknowledgments}

\bibliographystyle{apsrev4-1}
\bibliography{refs}

\end{document}